# Cell Painting Gallery: an open resource for image-based profiling

Erin Weisbart[1], Ankur Kumar[1], John Arevalo[1], Anne E. Carpenter[1], Beth A. Cimini[1], Shantanu Singh[1]

1. Broad Institute of MIT and Harvard, Cambridge MA, USA; Department: Imaging Platform

Image-based or morphological profiling is a rapidly expanding field wherein cells are "profiled" by extracting hundreds to thousands of unbiased, quantitative features from images of cells that have been perturbed by genetic or chemical perturbations. It is the least-expensive high-dimensional profiling technique to date, offers single cell resolution, and is successful in varied biological applications including early stage drug discovery (summarized in (Chandrasekaran et al., 2021)). The Cell Painting assay is the most popular imaged-based profiling assay wherein six small-molecule dyes label eight cellular compartments and thousands of measurements are made, describing quantitative traits such as size, shape, intensity, and texture within the nucleus, cytoplasm, and whole cell. First published in 2013 (Gustafsdottir et al., 2013), the standard protocol was updated in 2016 (Bray et al., 2016) and 2023 (Cimini et al., 2023).

The field of bioimaging, like most scientific fields, is in the middle of a sea change to make data FAIR (findable, accessible, interoperable, and reusable) (Schmied et al., 2023). In this vein, we have created the Cell Painting Gallery, a publicly available collection of Cell Painting datasets, with granular dataset descriptions and access instructions at https://github.com/broadinstitute/cellpainting-gallery. It is hosted by AWS on the Registry of Open Data (RODA). As of January 2024, the Cell Painting Gallery holds 656 terabytes (TB) of image and associated numerical data. It includes the largest publicly available Cell Painting dataset, in terms of perturbations tested (Joint Undertaking for Morphological Profiling or JUMP (Chandrasekaran et al., 2023)), along with many other canonical datasets using Cell Painting, close derivatives of Cell Painting (such as LipocyteProfiler (Laber et al., 2023) and Pooled Cell Painting (Ramezani et al., 2023)). Other sources of publicly available Cell Painting datasets include Recursion (Fay et al., 2023) (note that most metadata is currently blinded in their largest dataset) and IDR (Williams et al., 2017).

We aim for the Cell Painting Gallery data to be as useable as possible. To that end, we have implemented strict data and metadata guidelines, provided comprehensive download instructions, reprocessed old datasets, converted datasets to next-generation-file-formats, and worked with external organizations to make the Cell Painting Gallery browseable with their infrastructure, all detailed below.

All of the datasets within Cell Painting Gallery follow strict data and metadata organizational requirements, an important trait of FAIR data (Sarkans et al., 2021). This allows for images and numerical data to easily be downloaded/accessed separately, simplifying application-specific data download. Our data validator analyzes the contents of the gallery and generates customizable reports of attributes such as data integrity and dataset completion.

We provide detailed instructions to help researchers unfamiliar with accessing cloud-based data on AWS. Additionally, our lab has updated our Distributed-Science softwares (Weisbart & Cimini, 2023) (including Distributed-CellProfiler) to simplify performing compute in one AWS account while accessing images from another account (such as the Cell Painting Gallery). Similar updates to other image-based profiling tools that our lab maintains and contributes to are underway.

The field of morphological profiling has undergone major developments in the last 10 years, and like any scientific field, data that was state-of-the-art at publication may be a challenge to use and near impossible to align to datasets 10 years later. To bring important historical datasets up to date, we have reprocessed both cpg0012-wawer-bioactivecompoundprofiling and cpg0017-rohban-pathways, originally created in 2014 and 2017 respectively, using current state-of-the-field informatics (Cimini et al., 2023). Illustrating the utility of modernizing these datasets, the ~30,000 compounds in cpg0012-wawer-bioactivecompoundprofiling now have the same feature set as the ~136,000 chemical and genetic perturbations in cpg0016-jump, the JUMP CellPainting dataset (Chandrasekaran et al., 2023).

Proprietary image file formats are a major impediment to the reuse of both images and workflows developed to process and extract numeric data from images. OME-Zarr was developed as a next generation file format optimized for making bioimaging data FAIR and providing major improvements in cloud-based usage (Moore et al., 2021). It has rapidly grown an international community (Moore et al., 2023). We converted cpg0004-lincs to OME-Zarr using Distributed-OMEZarrCreator (Weisbart & Cimini, 2023) and plan to convert future datasets. Beyond improvements in image data accessibility, OME-Zarr conversion will improve interoperability with the Image Data Resource (IDR) to allow data hosted in the Cell Painting Gallery to be browsed through IDR after planned IDR server upgrades are complete.

The Cell Painting Gallery is continually expanding. Though the majority of the data, apart from JUMP, has been collected at the Broad Institute, we welcome data from other sources.

Acknowledgements

We thank Erin Chu and Nick Ragusa of AWS for assistance in getting the Cell Painting Gallery set up along with members of the Cimini and Carpenter-Singh labs for their feedback on this project and manuscript. This study was supported by Calico Life Sciences LLC, NIH P41 GM135019 (to BAC, AEC), NIH R35 GM122547 (to AEC), and CZI grant DAF2020-225720 (grant DOI https://doi.org/10.37921/977328pjvbca) from the Chan Zuckerberg Initiative DAF, an advised fund of Silicon Valley Community Foundation (funder DOI 10.13039/100014989) (to BAC). The funders had no role in study design, data collection and analysis, decision to publish or preparation of the manuscript.

# Code Availability

Cell Painting Gallery documentation is available at https://github.com/broadinstitute/cellpainting-gallery
Cell Painting Gallery data validator is available at https://github.com/broadinstitute/cpg

## Contributions

SS, BAC, and AEC conceived and supervised the project and revised the manuscript.
AK and JA wrote data validation scripts and revised the manuscript.
EW curated the data, wrote documentation and wrote the manuscript.

## Corresponding author

Correspondence to Shantanu Singh (shantanu@broadinstitute.org) and Erin Weisbart (eweisbar@broadinstitute.org)

# Ethics declarations

## Competing interests

The authors declare no competing interests.

| Dataset name | Description | Publication to cite | Objects | Total data size | Image data size | Numerical data size |
|---|---|---|---|---|---|---|
| cpg0000-jump-pilot | 300+ compounds and 160+ genes (CRISPR knockout and overexpression) profiled in A549 and U2OS cells, at two timepoints | Chandrasekaran, S. N. *et al.* Three million images and morphological profiles of cells treated with matched chemical and genetic perturbations. *bioRxiv* 2022.01.05.475090 (2022) doi:10.1101/2022.01.05.475090 | 5.2 M | 12.3 TB | 6.1 TB | 6.1 TB |
| cpg0001-cellpainting-protocol | 300+ compounds profiled in U2OS cells using several different modifications of the Cell Painting protocol | Cimini, B. A. *et al.* Optimizing the Cell Painting assay for image-based profiling. *Nat. Protoc.* (2023) doi:10.1038/s41596-023-00840-9 | 9.6 M | 40.3 TB | 18.7 TB | 21.6 TB |
| cpg0002-jump-scope | 300+ compounds profiled in U2OS using different microscopes and settings | Tromans-Coia, C. *et al.* Assessing the performance of the Cell Painting assay across different imaging systems. *Cytometry A* **103**, 915–926 (2023) | 2.6 M | 16.7 TB | 12.5 TB | 4.2 TB |
| cpg0003-rosetta | 28,000+ genes and compounds profiled in Cell Painting and L1000 gene expression | Haghighi, M., Caicedo, J. C., Cimini, B. A., Carpenter, A. E. & Singh, S. High-dimensional gene expression and morphology profiles of cells across 28,000 genetic and chemical perturbations. *Nat. Methods* **19**, 1550–1557 (2022) | 51 | 8.5 GB | 0 | 8.5 GB |
| cpg0004-lincs | 1,571 compounds across 6 doses in A549 cells | Way, G. P. *et al.* Morphology and gene expression profiling provide complementary information for mapping cell state. *Cell Syst* **13**, 911–923.e9 (2022) | 70.5 M | 65.7 TB | 61.9 TB | 3.8 TB |
| cpg0010-caie-drugre spons | MCF-7 breast cancer cells treated with 113 small molecules at eight concentrations. | Caie, P. D. *et al.* High-content phenotypic profiling of drug response signatures across distinct cancer cells. *Mol. Cancer Ther.* **9**, 1913–1926 (2010) | 1.1 M | 239.2 GB | 98.4 GB | 140.8 GB |
| cpg0011-lipocyteprofiler | Variety of lipocytes in different metabolic states and with genetic and drug perturbations | Laber, S. *et al.* Discovering cellular programs of intrinsic and extrinsic drivers of metabolic traits using LipocyteProfiler. *Cell Genomics* **3**, 100346 (2023) | 143 K | 1.2 TB | 1.2 TB | 16 MB |
| cpg0012-wawer-bioactivecompoundprofiling | 30,000 compound dataset in U2OS cells | Wawer, M. J. *et al.* Toward performance-diverse small-molecule libraries for cell-based phenotypic screening using multiplexed high-dimensional profiling. *Proc. Natl. Acad. Sci. U. S. A.* **111**, 10911–10916 (2014)<br><br>Bray, M.-A. *et al.* A dataset of images and morphological profiles of 30 000 small-molecule treatments using the Cell Painting assay. *Gigascience* **6**, 1–5 (2017) | 11 M | 10.7 TB | 3.1 TB | 7.6 TB |
| cpg0015-heterogeneity | 2,200+ compounds and 200+ genes profiles in U2OS cells | Rohban, M. H., Abbasi, H. S., Singh, S. & Carpenter, A. E. Capturing single-cell heterogeneity via data fusion improves image-based profiling. *Nat. Commun.* **10**, 2082 (2019) | 619 | 204 GB | 0 | 204 GB |
| cpg0016-jump | 116,000+ compounds and 16,000+ genes (CRISPR knockout and overexpression) profiled in over 1.5 billion U2OS cells. | Chandrasekaran, S. N. *et al.* JUMP Cell Painting dataset: morphological impact of 136,000 chemical and genetic perturbations. *bioRxiv* 2023.03.23.534023 (2023) doi:10.1101/2023.03.23.534023 | 115.3 M | 358.4 TB | 115.3 TB | 243 TB |
| cpg0017-rohban-pathways | 323 genes overexpressed in U2OS cells. Original images re-profiled in 2023 | Rohban, M. H. *et al.* Systematic morphological profiling of human gene and allele function via Cell Painting. *Elife* **6**, (2017) | 305 K | 321 GB | 189 GB | 132 GB |
| cpg0018-singh-seedseq | U2OS cells treated with each of 315 unique shRNA sequences | Singh, S. *et al.* Morphological Profiles of RNAi-Induced Gene Knockdown Are Highly Reproducible but Dominated by Seed Effects. *PLoS One* **10**, e0131370 (2015) | 138 K | 247.1 GB | 247.1 GB | 0 |
| cpg0019-moshkov-deepprofiler | 8.3 million single cells from 232 plates, across 488 treatments from 5 public datasets, used for learning representations | Moshkov, N. *et al.* Learning representations for image-based profiling of perturbations. Preprint at https://doi.org/10.1101/2022.08.12.503783 | 9.3 M | 522 GB | 482 GB | 40 GB |
| cpg0021-periscope | 30 million cells with 20,000 single-gene knockouts in pooled format. A549 cells and HeLa cells in two growth media | Ramezani, M. *et al.* A genome-wide atlas of human cell morphology. *bioRxiv* 2023.08.06.552164 (2023) doi:10.1101/2023.08.06.552164 | 7.1 M | 56.0 TB | 45.0 TB | 11.0 TB |
| cpg0022-cmqtl | 297 iPSC lines | Tegtmeyer, M. *et al.* High-dimensional phenotyping to define the genetic basis of cellular morphology. *Nat. Commun.* **15**, 347 (2024) | 702 K | 3.7 TB | 2.8 TB | 945 GB |
| cpg0028-kelley-resistance | Bortezomib resistant HCT116 clones | Kelley, M. E. *et al.* High-content microscopy reveals a morphological signature of bortezomib resistance. *Elife* **12**, (2023) | 1 M | 4.1 TB | 1.9 TB | 2.2 TB |
| cpg0030-gustafsdottir-cellpainting | U2OS cells treated with each of 1600 known bioactive compounds | Gustafsdottir, S. M. *et al.* Multiplex cytological profiling assay to measure diverse cellular states. *PLoS One* **8**, e80999 (2013) | 346 K | 234 GB | 234 GB | .3 GB |
| cpg0031-caicedo-cmvip | ORF over-expression of 596 alleles of 53 genes in A549 cells | Caicedo, J. C. *et al.* Cell Painting predicts impact of lung cancer variants. *Mol. Biol. Cell* **33**, ar49 (2022) | 553 K | 802 GB | 605 GB | 197 GB |

Table 1: Complete datasets available in the Cell Painting Gallery as of publication. Total data size (complete and in-progress datasets) as January 2024 is 656 TB.

## Works Cited

Bray, M.-A., Singh, S., Han, H., Davis, C. T., Borgeson, B., Hartland, C., Kost-Alimova, M., Gustafsdottir, S. M., Gibson, C. C., & Carpenter, A. E. (2016). Cell Painting, a high-content image-based assay for morphological profiling using multiplexed fluorescent dyes. *Nature Protocols*, *11*(9), 1757–1774. https://doi.org/10.1038/nprot.2016.105

Chandrasekaran, S. N., Ackerman, J., Alix, E., Michael Ando, D., Arevalo, J., Bennion, M., Boisseau, N., Borowa, A., Boyd, J. D., Brino, L., Byrne, P. J., Ceulemans, H., Ch'ng, C., Cimini, B. A., Clevert, D.-A., Deflaux, N., Doench, J. G., Dorval, T., Doyonnas, R., … Carpenter, A. E. (2023). JUMP Cell Painting dataset: morphological impact of 136,000 chemical and genetic perturbations. In *bioRxiv* (p. 2023.03.23.534023). https://doi.org/10.1101/2023.03.23.534023

Chandrasekaran, S. N., Ceulemans, H., Boyd, J. D., & Carpenter, A. E. (2021). Image-based profiling for drug discovery: due for a machine-learning upgrade? *Nature Reviews. Drug Discovery*, *20*(2), 145–159. https://doi.org/10.1038/s41573-020-00117-w

Cimini, B. A., Chandrasekaran, S. N., Kost-Alimova, M., Miller, L., Goodale, A., Fritchman, B., Byrne, P., Garg, S., Jamali, N., Logan, D. J., Concannon, J. B., Lardeau, C.-H., Mouchet, E., Singh, S., Shafqat Abbasi, H., Aspesi, P., Jr, Boyd, J. D., Gilbert, T., Gnutt, D., … Carpenter, A. E. (2023). Optimizing the Cell Painting assay for image-based profiling. *Nature Protocols*. https://doi.org/10.1038/s41596-023-00840-9

Fay, M. M., Kraus, O., Victors, M., Arumugam, L., Vuggumudi, K., Urbanik, J., Hansen, K., Celik, S., Cernek, N., Jagannathan, G., Christensen, J., Earnshaw, B. A., Haque, I. S., & Mabey, B. (2023). RxRx3: Phenomics Map of Biology. In *bioRxiv* (p. 2023.02.07.527350). https://doi.org/10.1101/2023.02.07.527350

Gustafsdottir, S. M., Ljosa, V., Sokolnicki, K. L., Anthony Wilson, J., Walpita, D., Kemp, M. M., Petri

Seiler, K., Carrel, H. A., Golub, T. R., Schreiber, S. L., Clemons, P. A., Carpenter, A. E., & Shamji, A. F. (2013). Multiplex cytological profiling assay to measure diverse cellular states. *PloS One*, *8*(12), e80999. https://doi.org/10.1371/journal.pone.0080999

Laber, S., Strobel, S., Mercader, J. M., Dashti, H., dos Santos, F. R. C., Kubitz, P., Jackson, M., Ainbinder, A., Honecker, J., Agrawal, S., Garborcauskas, G., Stirling, D. R., Leong, A., Figueroa, K., Sinnott-Armstrong, N., Kost-Alimova, M., Deodato, G., Harney, A., Way, G. P., … Claussnitzer, M. (2023). Discovering cellular programs of intrinsic and extrinsic drivers of metabolic traits using LipocyteProfiler. *Cell Genomics*, *3*(7), 100346. https://doi.org/10.1016/j.xgen.2023.100346

Moore, J., Allan, C., Besson, S., Burel, J.-M., Diel, E., Gault, D., Kozlowski, K., Lindner, D., Linkert, M., Manz, T., Moore, W., Pape, C., Tischer, C., & Swedlow, J. R. (2021). OME-NGFF: a next-generation file format for expanding bioimaging data-access strategies. *Nature Methods*, *18*(12), 1496–1498. https://doi.org/10.1038/s41592-021-01326-w

Moore, J., Basurto-Lozada, D., Besson, S., Bogovic, J., Bragantini, J., Brown, E. M., Burel, J.-M., Casas Moreno, X., de Medeiros, G., Diel, E. E., Gault, D., Ghosh, S. S., Gold, I., Halchenko, Y. O., Hartley, M., Horsfall, D., Keller, M. S., Kittisopikul, M., Kovacs, G., … Swedlow, J. R. (2023). OME-Zarr: a cloud-optimized bioimaging file format with international community support. *Histochemistry and Cell Biology*. https://doi.org/10.1007/s00418-023-02209-1

Ramezani, M., Bauman, J., Singh, A., Weisbart, E., Yong, J., Lozada, M., Way, G. P., Kavari, S. L., Diaz, C., Haghighi, M., Batista, T. M., Pérez-Schindler, J., Claussnitzer, M., Singh, S., Cimini, B. A., Blainey, P. C., Carpenter, A. E., Jan, C. H., & Neal, J. T. (2023). A genome-wide atlas of human cell morphology. In *bioRxiv* (p. 2023.08.06.552164). https://doi.org/10.1101/2023.08.06.552164

Sarkans, U., Chiu, W., Collinson, L., Darrow, M. C., Ellenberg, J., Grunwald, D., Hériché, J.-K., Iudin, A., Martins, G. G., Meehan, T., Narayan, K., Patwardhan, A., Russell, M. R. G., Saibil, H. R., Strambio-De-Castillia, C., Swedlow, J. R., Tischer, C., Uhlmann, V., Verkade, P., … Brazma, A.

(2021). REMBI: Recommended Metadata for Biological Images-enabling reuse of microscopy data in biology. *Nature Methods*, *18*(12), 1418–1422. https://doi.org/10.1038/s41592-021-01166-8

Schmied, C., Nelson, M. S., Avilov, S., Bakker, G.-J., Bertocchi, C., Bischof, J., Boehm, U., Brocher, J., Carvalho, M. T., Chiritescu, C., Christopher, J., Cimini, B. A., Conde-Sousa, E., Ebner, M., Ecker, R., Eliceiri, K., Fernandez-Rodriguez, J., Gaudreault, N., Gelman, L., … Jambor, H. K. (2023). Community-developed checklists for publishing images and image analyses. *Nature Methods*. https://doi.org/10.1038/s41592-023-01987-9

Weisbart, E., & Cimini, B. A. (2023). Distributed-Something: scripts to leverage AWS storage and computing for distributed workflows at scale. *Nature Methods*. https://doi.org/10.1038/s41592-023-01918-8

Williams, E., Moore, J., Li, S. W., Rustici, G., Tarkowska, A., Chessel, A., Leo, S., Antal, B., Ferguson, R. K., Sarkans, U., Brazma, A., Salas, R. E. C., & Swedlow, J. R. (2017). The Image Data Resource: A Bioimage Data Integration and Publication Platform. *Nature Methods*, *14*(8), 775–781. https://doi.org/10.1038/nmeth.4326